\documentclass[preprint,12pt]{elsarticle}

\usepackage[T1]{fontenc}
\usepackage[latin1]{inputenc}
\usepackage{graphicx}
\usepackage{amssymb}
\usepackage{amsmath}
\usepackage{dcolumn}
\usepackage{bm}
\usepackage{color}
\usepackage{multicol}
\usepackage{algorithm}
\usepackage{multirow}
\usepackage{indentfirst}
\usepackage{graphicx}
\usepackage{subfigure}
\usepackage{epstopdf}
\usepackage{diagbox}

\newcommand{\w}{\mathbf{w}}

\graphicspath{ {./DirectedFigures/} }

\newcommand{\eproof}{\hfill\rule{2mm}{2mm}}

\newtheorem{theorem}{Theorem}
\newtheorem{definition}{Definition}
\newtheorem{lemma}{Lemma}

\newtheorem{remark}{Remark}

\journal{ISA Transactions}

\begin{document}

\begin{frontmatter}
 


\title{Structural Consensus in Networks with Directed Topologies and Its Cryptographic Implementation}



 \author{Wentuo Fang}
 \ead{Wentuo.fang@uon.edu.au}

 \author{Zhiyong Chen\corref{cor1}}
 \ead{zhiyong.chen@newcastle.edu.au}
\cortext[cor1]{Tel: +61 2 4921 6352, Fax: +61 2 4921 6993}
 \address{School of Engineering, The University of Newcastle, Callaghan, NSW 2308, Australia}

\author{Mohsen Zamani}
\ead{mohsen.zamani@newcastle.edu.au}
\address{Department of Medical Physics and Engineering, Shiraz University of Medical Sciences, 
Shiraz, Zand, Iran}

\begin{abstract}
The existing cryptosystem based approaches for privacy-preserving consensus of networked systems are usually limited to those with undirected topologies. This paper proposes a new privacy-preserving algorithm for networked systems with directed topologies to reach confidential consensus.  As a prerequisite for applying the algorithm, a structural consensus problem is formulated and the solvability conditions are discussed for an explicitly constructed controller. The controller is then implemented with encryption to achieve consensus while avoiding individual's information leakage to external eavesdroppers and/or malicious internal neighbors.  

\end{abstract}



\end{frontmatter}

 \newpage

\section{Introduction}

 Consensus of multi-agent systems (MASs) is one of the most commonly used protocols in 
 networked control systems  with a wide range of applications. 
It aims to make all agents to reach an agreement.
Some results for consensus of  first order agents can be found in many early papers. For example, 
analysis of Laplacian was used to formulate the convergence conditions in \cite{olfati-saber_consensus_2004}
and a convergence proof for discrete-time systems based on a theoretical property of infinite products of stochastic matrices
was presented in \cite{jadbabaie_coordination_2003}.  More works on consensus of higher order and nonlinear systems include \cite{xie_consensus_2012, lin_consensus_2009, yu_necessary_2010, yu_distributed_2011} and references therein.

Network communication is an essential factor in a consensus problem.  Over a network,  the information, which usually refers to agents' states, is transmitted and shared. However, passing messages through a common network lead to risk of information leakage or tampering. This has motivated significant amount 
of researches on maintaining security and privacy of individuals' information  within a network. It is worth noting that network topology plays an important role in  many  performance characteristics of a network such as consensus value, convergence rate and its information privacy. 
For example,  a linear system with an undirected topology achieves average consensus \cite{zhu_discrete-time_2010, kingston_discrete-time_2006}, while that with a directed leader-following topology leads to 
consensus on the leader's state.

During a consensus process, information exchange is critical while agents also intend to keep their states confidential to both external eavesdroppers and/or other internal agents within the network.  
In order to prevent information disclosure or tampering attack, different models and techniques for network security and privacy have been widely studied.
Models involving adversarial agents attempting to make  a system unstable were studied in \cite{leblanc_consensus_2011, leblanc_consensus_2012}. 
A connectivity-broken attack model was introduced in \cite{feng_distributed_2016}.
Differentially private iterative consensus was studied in different settings; see, e.g., \cite{manitara_privacy-preserving_2013, mo_privacy_2017} 
for privacy-preserving average consensus and  \cite{duan_privacy_2015} for maximum consensus.


Cryptosystem is another important tool for information security \cite{ruan_secure_2017, fang_secure_2018}. It has been used in various privacy-preserving algorithms. 
For instance, a Paillier encryption based cloud-computing algorithm was introduced in \cite{krutz_cloud_2010, farokhi_secure_2016}. The cloud server  is a trusted third-party which receives encrypted data from agents and 
executes computation. 
A Paillier encrypted observer-based control paradigm was proposed in \cite{sadeghikhorami2020secure}
for enhancement of cybersecurity in networked control systems.  
Paillier encryption was  also exploited in distributed networks \cite{ruan_secure_2017,ruan_secure_2019-1}.
Introducing cryptosystem to MASs may prevent agents' states from being detected by eavesdroppers.
However, due to the difficulties of applying privacy-preserving algorithms while guaranteeing system stability, so far most of the privacy-preserving algorithms are limited to linear systems with undirected network topologies which lead to average consensus; see, e.g., \cite{manitara_privacy-preserving_2013, mo_privacy_2017,he_privacy-preserving_2019, liu_secure_2017}.
The  research work proposed in this paper follows this line of research and focuses on directed network topologies.

It should be noted that there exist  substantial differences in designing privacy-preserving algorithms for MASs under undirected and directed topologies. 
On one hand,  to allow privacy-preserving algorithms, we must  introduce time-varying weights that are capable of carrying proper randomness. 
This however induces the technical challenge in maintaining consensus with time-varying weights.
To formulate the consensus behavior in this scenario, we introduce the concept of structural consensus in this paper. 
It basically requires achievement of consensus with arbitrary variation (within a specified range) of nonzero weights while 
the zero weights are always kept. 
The concept is borrowed from the  structural controllability studied in \cite{lin1974structural,mehrabadi_structural_2019}
and the references therein.
Usually, it is impossible to achieve  consensus in a network equipped with
directed time-varying weights which does not ensure a constant left eigenvector 
associated with the zero eigenvalue for the asymmetric time-varying Laplacian.

 On the other hand, in undirected networks, a pair of neighbors use the same coupling weights, which brings more risk of individuals' states leakage. For instance, information disclosure occurs when an agent knows it is the only neighbor of another agent; see, e.g., \cite{ruan_secure_2017,fang_secure_2018}. Whereas in directed networks, this risk is lower since the coupling weights become asymmetric.  
It motives us to introduces a new Paillier encryption based privacy-preserving consensus algorithm for systems under directed leader-following topologies. We propose a confidential communication algorithm and the consensus conditions, under which all agents can exchange information with neighbors and eventually reach the agreement without leakage of states to either neighbors or eavesdroppers. 
By applying the privacy-preserving algorithm, some uncertainties are brought into the nonzero coupling weights of system's network topology. 
In particular,  the sufficient conditions for both first order and second order MASs to reach structural consensus are presented and the differences are discussed.

The rest of this paper is organized as follows.  The system dynamics and problem formulation are presented in Section~\ref{sec-problem_formulation}. In Section~\ref{sec-consensus}, the conditions for structural consensus under the proposed paradigm are provided. 
Section~\ref{sec-cryptography} presents the cryptographic communication algorithms for both first order and second order MASs. 
Section~\ref{sec-simulation} gives a numerical example. Finally, Section \ref{sec-conclusion} concludes the paper with some remarks.

\section{Problem Formulation}\label{sec-problem_formulation}

In this section, we introduce the network topology, system dynamics and formulate the
structural consensus problem and its cryptographic implementation
with the attack model under study.
Specifically, we consider a network  of $N\geq 2$ agents with a directed leader-following topology.

Let $G$ ($\mathcal{V}$, $\mathcal{E}$) denote the graph  of a network of agents where the set of nodes and edges are presented by $\mathcal{V}$ and $\mathcal{E}$, respectively. 
Each element of $\mathcal{V}=\{1,\cdots, N\}$  represents one node (agent)
and each element $(i,j) \in \mathcal{E}$ represents the edge from $i$ to $j$.   A directed path is a sequence of nodes $i_1,i_2, \cdots i_r$ such that $(i_1,i_2), (i_2,i_3),  \cdots, (i_{r-1}, i_r)\in \mathcal{E}$.
Node $i$ is said to be connected to node $j$ if there is a directed path from $i$ to $j$. A set of nodes which are connected to $i$ is defined as its neighbors, i.e., $\mathbb{N}_i= \{ j \in \mathcal{V} \; |\; (j,i) \in\mathcal{E} \}$. 
A connected graph without cycles is called a tree. A directed spanning tree of graph $G$ is a tree which includes all nodes in $G$ and there exists a directed path from root to any other nodes. The root of a spanning tree is called a leader.
The adjacency matrix $A$ with its $(i,j)$-entry $a_{ij}$ is defined as follows: 
$a_{ij}>0$ if $(j,i)\in \mathcal{E}$; and  $a_{ij}=0$ otherwise.  The Laplacian matrix $L$ 
with its $(i,j)$-entry $l_{ij}$
is defined as $l_{ij}=-a_{ij},i\neq j$, and 
$l_{ii}=\sum_{j=1,j\neq i}^{N}a_{ij}$. 
 
 \medskip
 
A simple MAS model of first order dynamics is represented by the following discrete-time difference equations
\begin{align}\label{equ-model_1}
x_i[k+1]=x_i[k] + u_i[k],\; i=1,2,\dots, N, 
\end{align}
where $k=0,1,\dots$ represents the time sequence throughout the paper. 
In this model, $x_i \in {\mathbb R}$ is the agent state and the control input is given by 
 \begin{align}\label{equ-model-input_1}
u_i[k]=\epsilon \sum_{j\in \mathbb{N}_i} a_{ij}[k](x_j[k]-x_i[k]),\;
  i=1,2,\dots, N,
\end{align}
 where $\epsilon>0$ is the iteration step size and $a_{ij}[k]$'s are the time-varying coupling weights defined as the $(i,j)$-entry of the adjacency matrix $A[k]$  at the time $k$ with 
$a_{ij}[k]\geq 0$ and $a_{ii}[k] =0$. 
In particular, $a_{ij}[k] > 0$ for $j\in \mathbb{N}_i$ and $a_{ij}[k] = 0$  otherwise.

The closed-loop system composed of \eqref{equ-model_1} and \eqref{equ-model-input_1}
can be rewritten in an aggregated form  as follows,
\begin{align}\label{equ-SystemModel_G}
x[k+1]=F[k] x[k]
\end{align}
where
\begin{align} \label{F1}
x=\left[x_1, x_2,\dots, x_N\right]^\mathsf{T},\; F[k]=I-\epsilon L[k].
\end{align} 

\begin{remark}
	In this paper, we consider a scenario where the system network represented by the
adjacency matrix $A[k]$  is directed in the sense of exploiting information, whereas the communication channels between agents are bidirectional. The reason for using bidirectional communication comes from the requirements of applying public-key cryptosystem, which allows agents to acquire a needed value without knowing the process variables in computing.
\end{remark}

An MAS model of second order dynamics is represented by 
 \begin{align}\label{equ-model_2}
p_i[k+1]&=p_i[k]+v_i[k], \nonumber\\
v_i[k+1]&=v_i[k]+u_i[k], \; i=1,2\dots,N,
\end{align}
where $p_i, v_i\in \mathbb{R}$ are the states of the agent $i$ and can be regarded as the position and velocity,  respectively.    
The following control law drives the group of agents  in \eqref{equ-model_2} toward reaching
consensus asymptotically
\begin{align}\label{equ-model-input_2}
u_i[k]=&\gamma_1\sum_{j\in \mathbb{N}_i}a_{ij}[k](p_j[k]-p_i[k]) \nonumber\\
&+\gamma_2\sum_{j\in \mathbb{N}_i}a_{ij}[k](v_j[k]-v_i[k]),  \; i=1,2\dots,N,
\end{align}
for  two positive coefficients $\gamma_1$ and $\gamma_2$ satisfying a certain conditions.
Again, the closed-loop system can be put in the form \eqref{equ-SystemModel_G} with
\begin{align} \label{F2}
p&=\left [ p_1, p_2, \cdots, p_N \right]^\mathsf{T},\;
v=\left [ v_1, v_2, \cdots, v_N \right]^\mathsf{T}, 
\nonumber\\
x&=\left[ p^\mathsf{T},v^\mathsf{T}\right]^\mathsf{T}, \;
F[k] =
\begin{pmatrix}
I_N &  I_N\\
-\gamma_1 L[k] & I_N-\gamma_2 L[k]
\end{pmatrix}.
\end{align}
 Next, we will formulate the main problem studied in this paper. 
For this purpose, we specify a constant $\delta$ as follows,
\begin{align}
\label{delta}
0<\delta<a_{ij}[0], \; \forall 1\leq i,j \leq N, a_{ij}[0] > 0,
\end{align} 
which denotes a lower bound of the nonzero coupling weights at the initial time.

\begin{definition}\label{de-StructuralConsensus}
The MAS \eqref{equ-SystemModel_G} is said to achieve {\it structural consensus} if 
there exists $\sigma[k] \in \mathbb{R}$ (first order dynamics)
or $\sigma[k] \in \mathbb{R}^2$ (second order dynamics) such that
\begin{align} \label{consensus}
\lim_{k\rightarrow\infty} (x[k] -\sigma[k] \otimes \mathbf{1}_N) =0,
\end{align} 
for any \begin{align}\label{aij}
a_{ij}[k] \left\{
\begin{array}{ll}
\in [a_{ij}[0]-\delta,a_{ij}[0]+\delta], &  a_{ij}[0] >0 \\
=0, & a_{ij}[0] =0
\end{array}
\right. .
\end{align}
Here, $\mathbf{1}_N$ is the $N$-dimensional vector whose entries are 1. 
\end{definition}

\begin{remark}
\label{rmk-eigenvector_LF}
The network weights are time-varying in the setting. 
  Note that once the iteration step size $\epsilon$ in \eqref{F1} is decided, the stability of first-order systems can also be proved using the property of products of stochastic matrices. However, the results cannot be extended to second-order systems due to the fact that $F[k]$ in \eqref{F2} is no longer stochastic.
For second order systems,  it is usually impossible to achieve consensus in a network equipped with
directed time-varying weights which does not ensure a constant left eigenvector 
associated with the zero eigenvalue for the asymmetric time-varying Laplacian. 
Therefore, we consider the special leader-following topology whose Laplacian matrix $L[k]$ 
always attains an eigenvalue $0$ associated with a right eigenvector $\mathbf{1}_N$ 
and a constant left eigenvector $\w=(0,\cdots, 0, 1,0,\dots,0)^{\mathsf{T}}$ with 
the entry $1$  corresponding to the leader.  
\end{remark}

 As explained earlier, our main objective is to establish privacy-preserving consensus strategy for a network of agents with both  first order and second order dynamics under a directed topology. To this end, the first task of this paper is to find 
the conditions for $\epsilon$  (first order dynamics)
or $\gamma_1$ and $\gamma_2$ (second order dynamics)
under which the structural consensus is achieved.

 Then the second task is to consider agent privacy when implementing the consensus algorithm. 
In particular,  we consider disclosure attack by malicious neighbors and/or eavesdroppers. 
The objective is to keep each agent's information private to itself as defined below.
In other words, we do not consider false signal injection attack, which can be detected by, e.g., applying digital signature.
\begin{definition}\label{def-PrivacyPreserved}
	An agent's privacy is said to be preserved if its initial state remains unknown to others during the whole consensus process. 
\end{definition}

 In the present scenario, the adversaries are the neighbors or eavesdroppers that receive and collect communicating messages in order to compute or estimate the agent's initial states.
If the initial states are disclosed, an agent's trajectory can be reconstructed easily since system dynamics, i.e., \eqref{equ-model_1} or \eqref{equ-model_2}, are assumed to be public knowledge.  
 To achieve this  task, we use cryptographic implementation (using Paillier encryption) 
of the consensus algorithm as well
as keeping the network weights $a_{ij}[k]$ secret with the agent who generates them.
It  then becomes apparent that structural consensus  as the  first task is the prerequisite
which allows randomness in generating time-varying $a_{ij}[k]$  required for establishing the second task.

\section{Structural Consensus}\label{sec-consensus}

In this section, we aim to give the explicit conditions 
for $\epsilon$  (first order dynamics)
or $\gamma_1$ and $\gamma_2$ (second order dynamics)
under which structural consensus is achieved,  thus achieving the first  task.

\subsection{First order Dynamics}\label{ssec-convergence_analysis}

\begin{theorem}\label{lem-stable}
Consider the MAS \eqref{equ-SystemModel_G} with \eqref{F1}
in a network equipped with a directed leader-following topology containing a spanning tree. 
For a specific $\delta$ satisfying \eqref{delta} and  $a_{ij}[k]$ satisfying \eqref{aij}, the MAS achieves structural consensus if 
the parameter $\epsilon$ satisfies
	\begin{equation}\label{con-stable}
	\epsilon<\frac{1}{\max \limits_{1\le i\le N}(\|\alpha_i\|_1+\delta\|\alpha_i\|_0)},
	\end{equation}
	where $\alpha_i$ is the $i$-th  row of $A[0]$, $\|\mathbf{\cdot}\|_1$ and $\|\mathbf{\cdot}\|_0$ denote $1$-norm and $0$-norm, respectively.
\end{theorem}

\noindent{\it Proof:}
In the first order system under a directed leader-following topology containing a spanning tree, 
the leader is a special agent whose dynamic as in \eqref{equ-model_1} has control input $u_1[k]=0$.   
From the structure analysis of the Laplacian matrix $L[k]$ of such a 
topology, it has an eigenvalue $0$ associated with a left eigenvector $\w$ and a right eigenvector $\mathbf{1}_{N}$. Then, the iteration matrix $F[k]$ has an eigenvalue $1$ associated with the same eigenvectors. 


From (\ref{aij}),  the diagonal elements of $L$ satisfy, for $i=1,2,\dots,N$,
	\begin{equation}\label{equ-stable_proof_diagonal_Laplacian}
	|l_{ii}[k]|\le|l_{ii}[0]|+\delta\|\alpha_i\|_0.
	\end{equation}	
	Since $|l_{ii}[0]|=\|\alpha_i\|_1$, the inequality \eqref{equ-stable_proof_diagonal_Laplacian} together with \eqref{con-stable}, implies 
	\begin{equation}\label{equ-stable_condition_1}
		|\epsilon l_{ii}[k]|<1.		
	\end{equation} 
	According to the Gershgorin Circle Theorem, every eigenvalue $\lambda$ of the matrix $\epsilon L[k]$ lies within at least one of the Gershgorin discs 
	\begin{equation} D(\epsilon l_{ii}[k], \epsilon\sum\limits_{j\ne i}|l_{ij}[k]|) =
	D(\epsilon l_{ii}[k],\epsilon l_{ii}[k]).\end{equation}
Here, $D(M,R)$ represents a circle closed set centered at $M$ with radius $R$. 
From \eqref{equ-stable_condition_1}, 
every eigenvalue $\lambda$ of the matrix $\epsilon L[k]$ lies within the interior of the set $D(1,1)$
except one eigenvalue $0$.
Furthermore, the iteration matrix $F[k]$ has one eigenvalue $1$ and all the others within the 
unit circle.

	Next, we will prove that the system  \eqref{equ-SystemModel_G} of such an iteration matrix $F[k]$ converges to the leader's state.	 	
	Let $J$ be the Jordan canonical form of $F[0]$. The following transformation can be defined
	\begin{equation}\label{equ-stable-jordan}
	F[0]=PJP^{-1},
	\end{equation}
	where $J=
	\begin{bmatrix}
	1 & \mathbf{0}_{N-1}^\mathsf{T}\\
	\mathbf{0}_{N-1} & \tilde{J}
	\end{bmatrix},
	$
	$P=
	\begin{bmatrix}
	 \mathbf{1}_{N} & M_{N\times(N-1)}
	\end{bmatrix}
	$
	and $P^{-1}=
	\begin{bmatrix}
	\w^\mathsf{T}  \\
	W_{(N-1)\times N}
	\end{bmatrix}.
	$
	Note that all the eigenvalues of $\tilde{J}$ lie within the unit circle.
	For $k=1,2,\dots$, the similarity transformation on  $F[k]$ gives
	\begin{equation}\label{equ-stable-jordan_not}
	F[k]=P Q[k] P^{-1}
	\end{equation} for $Q[k]=
	\begin{bmatrix}
	1 & \mathbf{0}_{N-1}^\mathsf{T}\\
	\mathbf{0}_{N-1} & \tilde{Q}[k]
	\end{bmatrix}.$ As $Q[k]$ is a similar matrix of $F[k]$, $\tilde{Q}[k]$ is a Schur matrix with all eigenvalues within the unit circle.
	
	At time $k$, the state $x[k]$ can be calculated by
	\begin{align*}
	x[k] = (\prod^{0}_{\tau=k-1}F[\tau]) x[0]  	= P
	\begin{bmatrix}
	1 & \mathbf{0}_{N-1}^\mathsf{T}\\
	\mathbf{0}_{N-1} & \prod\limits^{0}_{\tau=k-1} \tilde{Q}[\tau]
	\end{bmatrix}
	P^{-1} x[0].
	\end{align*}
	Since  $\tilde{Q}[k],  k=0,1,2,\dots$, are Schur matrices whose eigenvalues 
have norms upper bounded by a constant less than 1,  one has
	$\lim_{k\to\infty}\prod\limits^{0}_{\tau=k-1} \tilde{Q}[\tau] =\mathbf{0}_{(N-1)\times(N-1)}$, and hence 	
	\begin{align}
	\lim\limits_{k\to \infty} x [k]
	&= P
	\begin{bmatrix}
	1 & \mathbf{0}_{N-1}^\mathsf{T}\\
	\mathbf{0}_{N-1} & \mathbf{0}_{(N-1)\times(N-1)}
	\end{bmatrix}
	P^{-1} x [0] \nonumber \\
	&= (\w^\mathsf{T} x[0]) \mathbf{1}_{N}.
	\end{align}
Therefore, structural consensus is achieved in the sense of \eqref{consensus} with $\sigma[k] =\w^\mathsf{T} x[0]$.
\eproof

\begin{remark}
An MAS of first order dynamics has the matrix $F[k]$ defined in 
\eqref{F1}. In order to reach consensus, all the eigenvalues of $F[k]$ must be within the unit circle 
in complex plane, except for the eigenvalue $1$. This requires $\epsilon L[k]$'s eigenvalues to be within the circle centered at $(-1,0)$ with radius of $1$, which can be guaranteed by choosing a sufficiently large $\epsilon$ according to the Gershgorin Circle Theorem, because all the eigenvalues of $L[k]$ stay in a disk which is in the right half complex plane and is tangent to the y-axis. 
A larger $\epsilon$ makes Gershgorin disk smaller.
	\end{remark}
 
 \subsection{Second order Dynamics}\label{ssec-convergence_analysis_2}

We first give two technical lemmas. The main theorem follows.
Denote the zero eigenvalue of the Laplacian $L[k]$ by $\mu_1[k]=0$. The other $N-1$ eigenvalues, 
denoted by $\mu_i [k], \; i=2,\cdots, N$,
have the following property. 

 \begin{lemma} \label{lemma:eigenvalue}
For a specific $\delta$ satisfying \eqref{delta} and $a_{ij}[k]$ satisfying \eqref{aij}, 
there exist $\pi/2 > \theta_{\max}>0$ and $r_{\min}, r_{\max} >0$, such that the nonzero eigenvalues of $L[k]$ are 
\begin{align}
\mu_i [k] = r_i[k] e^{{\rm j} \theta_i[k] }, \; i=2,\cdots, N
\end{align}
for
\begin{align}\label{thetarb}
\theta_i[k] \in [-\theta_{\max} ,\; \theta_{\max}],\; 
r_i [k]\in [r_{\min} ,\; r_{\max}],
\end{align}
if the directed leader-following topology contains a spanning tree.
\end{lemma}

\noindent{\it Proof:}
 The Laplacian matrix $L[k]$ has $N-1$ eigenvalues of positive real parts as
the topology contains a spanning tree (see, e.g., Theorem~1 in \cite{olfati-saber_consensus_2004}), that is, 
$-\pi/2 < \theta_i[k] <\pi/2,  \; r_i [k] >0,\;i=1,2,\dots N$.
The eigenvalues continuously depend on the parameters $a_{ij}[k],\; i,j=1,2,\dots N$,
of $L[k]$. Therefore,  with the change to $a_{ij}[k]$ in a compact set,  i.e., $a_{ij}[k] \in [a_{ij}[0]-\delta, a_{ij}[0]+\delta]$,
the eigenvalues are  also in a compact set. 
As a result,  there exist $\pi/2 > \theta_{\max}>0$ and $r_{\min}, r_{\max} >0$
to make boundaries of $\theta_i[k]$ and $r_i [k]$ in the sense of  \eqref{thetarb}. The proof is thus completed.
\eproof

\begin{lemma}\label{lemma:parameter}
	Consider $\varrho <1$ and  $\kappa >0$ satisfying the following inequalities  \begin{align}
\frac{ 1    - \sqrt{  1 -    \varrho   }  }   { r_{\min} }
<  \kappa 
< \frac{ 1   +   \sqrt{  1 -    \varrho   }    } { r_{\max} }\cos \theta_{\max}  \label{kappa}
\end{align}
for $\theta_{\max}$, $r_{\min}$ and $r_{\max} $ given in Lemma~\ref{lemma:eigenvalue}.
Let $\rho >1$ be the solution to
\begin{align*}
\frac{ 2\rho-1   }{( \rho-1)^2   }  &=    \varrho \cot \theta_{\max}.
\end{align*}
Then, the matrix $F[k]$ in  \eqref{F2} with 
\begin{align}
   \gamma_1  =\frac{2 \kappa }{2 \rho-1} ,\;
  \gamma_2 = \rho \gamma_1,
\end{align}
has $2(N-1)$ eigenvalues  whose norms are upper bounded by a constant less than 1
and an eigenvalue $1$ of algebraic multiplicity two.  
 \end{lemma}

\noindent{\it Proof:}
Consider \eqref{equ-SystemModel_G} with $F[k]$ defined in \eqref{F2}.  Given that the Laplacian  $L[k]$ has one $0$ eigenvalue associated with eigenvector $\mathbf{1}_N$ and all the others $N-1$ positive, one can easily conclude that $F[k]$ has an eigenvalue $1$ of algebraic multiplicity two. 
The corresponding generalized right eigenvectors are $\mathbf{r}_{F1}=(\mathbf{1}_N^{\mathsf{T}},\mathbf{0}_N^{\mathsf{T}})^{\mathsf{T}}$ and $\mathbf{r}_{F2}=(\mathbf{0}_N^{\mathsf{T}},\mathbf{1}_N^{\mathsf{T}})^{\mathsf{T}}$ and the generalized left eigenvectors 
	$\w_{F1}=(\mathbf{0}_N^\mathsf{T},\w^\mathsf{T})^\mathsf{T}$ and $\w_{F2}=( \w^\mathsf{T}, \mathbf{0}_N^\mathsf{T} )^\mathsf{T}$.

The existence of  $\varrho <1$ and  $\kappa >0$ satisfying \eqref{kappa}
is guaranteed by the  fact that there exists a sufficiently small $\varrho$ satisfying
  \begin{align*}
\frac{ 1    - \sqrt{  1 -    \varrho   }    }   {1    +  \sqrt{  1 -    \varrho   } }
< \frac{   r_{\min}\cos \theta_{\max}   } {r_{\max} }.
\end{align*}
The solutions to the following second polynomial equation of $r_i$,
\begin{align*}
\gamma_1 (2 \rho-1) r_i^2 -4 r_i \cos\theta_i + \frac{4 \sin^2\theta_i }{( \rho-1)^2 \gamma_1} =0,
\end{align*}
are
\begin{align} \label{curve}
 r_i^*(\theta_i) =  \frac{2 \cos\theta_i \pm 2 \sqrt{  \cos^2\theta_i 
 -   \frac{  (2 \rho-1)  \sin^2\theta_i }{( \rho-1)^2 }}} {\gamma_1 (2 \rho-1)}.
\end{align}
From
\begin{align*}
 \frac{  (2\rho-1)   }{( \rho-1)^2 \varrho } = \cot \theta_{\max} \leq \frac{  \cos^2\theta_i }{  \sin^2\theta_i },
\end{align*}
one has
\begin{align*}
& \frac{2 \cos\theta_i  + 2 \sqrt{  \cos^2\theta_i 
 -     \frac{ (2 \rho-1) \sin^2\theta_i }{( \rho-1)^2 }}}{\gamma_1 (2 \rho-1)}  \\
  \geq & \frac{2( 1    +  \sqrt{  1 -    \varrho   } ) \cos \theta_i  }{\gamma_1 (2 \rho-1)}
  \geq  \frac{( 1    +  \sqrt{  1 -    \varrho   } ) \cos \theta_{\max}  }{\kappa} > r_{\max}.
\end{align*}
and
\begin{align*}
& \frac{2 \cos\theta_i  - 2 \sqrt{  \cos^2\theta_i 
 -     \frac{ (2 \rho-1) \sin^2\theta_i }{( \rho-1)^2 }}}{\gamma_1 (2 \rho-1)}  \\
  \leq & \frac{2( 1    -  \sqrt{  1 -    \varrho   } ) \cos \theta_i  }{\gamma_1 (2 \rho-1)}
  \leq  \frac{ 1    -  \sqrt{  1 -    \varrho   }    }{\kappa}  < r_{\min}.
\end{align*}
As a result, for any $r_i \in [r_{\min} ,\; r_{\max}]$, 
\begin{align*}
\gamma_1 (2 \rho-1) r_i^2 -4 r_i \cos\theta_i + \frac{4 \sin^2\theta_i }{( \rho-1)^2 \gamma_1} < 0,
\end{align*}
that is equivalent to
\begin{align*}
 (2 \gamma_2-\gamma_1) r_i^2 -4 r_i \cos\theta_i + \frac{4  \gamma_1 \sin^2\theta_i }{( \gamma_2- \gamma_1)^2  } < 0,
\end{align*}
or
\begin{align}
 (2 \gamma_2-\gamma_1) |\mu_i [k]|^2 -4  \Re(\mu_i[k]) + \frac{4  \gamma_1\Im^2(\mu_i[k]) }{( \gamma_2- \gamma_1)^2  |\mu_i [k]|^2} < 0.
 \label{gammacon1}
\end{align}
Next, one has
\begin{align*}
\gamma_1-2\gamma_2 &= \frac{2 \kappa (1-2\rho) }{2 \rho-1}=-2\kappa
\\ &>-2\frac{ 1   +   \sqrt{  1 -    \varrho   }    } { r_{\max} }\cos \theta_{\max}
>\frac{-4   \cos\theta_i}{r_i},
\end{align*}
that is equivalent to
\begin{align}
\gamma_1-2\gamma_2>\frac{-4\Re(\mu_i[k])}{|\mu_i[k]|^2}.
\label{gammacon2}
\end{align}
Obviously, one has 
\begin{align}
\gamma_2>\gamma_1>0\label{gammacon3}
\end{align} due to $\rho >0$.

Theorem~1 and Theorem~2 of \cite{xie_consensus_2012} 
reveal the connection between the spectrum of $F[k]$ in \eqref{F2} and the parameters $\gamma_1$ and $\gamma_2$. 
In particular, if
 $\gamma_1$ and $\gamma_2$ satisfy \eqref{gammacon1}, \eqref{gammacon2}, and 
	\eqref{gammacon3} for all the nonzero eigenvalues $\mu_i[k]$ of $L[k]$, then $F[k]$ has $2(N-1)$ eigenvalues inside the unit cycle (whose norms are less than one) 	and an eigenvalue $1$ of algebraic multiplicity two.
Moreover, as $\mu_i [k], \;i=2,\cdots, N$, are valued in a compact set 
for all $k$, so are the $2(N-1)$ eigenvalues of $F[k]$. As a results, 
the norms of these $2(N-1)$ eigenvalues  are upper bounded by a constant less than 1.
This completes the proof. 
\eproof

By using these two technical lemmas, we can prove the following theorem.

\begin{theorem} 
 Consider the MAS \eqref{equ-SystemModel_G} with \eqref{F2}
in a network equipped with a directed leader-following topology containing a spanning tree. 
For a specific $\delta$ satisfying \eqref{delta} and $a_{ij}[k]$ satisfying \eqref{aij}, 
the MAS achieves structural consensus if 
the parameters $\gamma_1$ and $\gamma_2$ are selected in Lemmas~\ref{lemma:eigenvalue} and 
\ref{lemma:parameter}.
\end{theorem}

\noindent{\it Proof:} From Lemmas~\ref{lemma:eigenvalue} and 
\ref{lemma:parameter},   the matrix $F[k]$ 
has $2(N-1)$ eigenvalues  whose norms are upper bounded by a constant less than 1
and an eigenvalue $1$ of algebraic multiplicity two.   Let us introduce a nonsingular matrix:
\begin{align*}
P=
	\begin{bmatrix}
	\mathbf{r}_{F1} & \mathbf{r}_{F2}
			& M_{2N\times(2N-2)}
	\end{bmatrix}, \;
	P^{-1}=
	\begin{bmatrix}
	\w^\mathsf{T}_{F1} \\
	\w^\mathsf{T}_{F2} \\
	W_{(2N-2)\times 2N}
	\end{bmatrix}.
\end{align*}
It gives  the Jordan canonical form of $F[k]$ as follows
\begin{align} \label{equ-Jordanform-F}
	Q[k]= P^{-1} F[k] P   
	=
	\begin{bmatrix}
	\begin{matrix}
	1 & 1\\
	0 & 1
	\end{matrix} &
	\mathbf{0}
	_{2\times (2N-2)}
	\\  \mathbf{0}
	_{(2N-2)\times 2}
	& \tilde{Q}[k] 
	\end{bmatrix}
\end{align}
where $\tilde{Q}[k]=W F[k] M$ is a Schur matrix  whose eigenvalues have norms upper bounded by a constant less than 1.
Next, we can calculate $x[k]$ by
\begin{align*}
	x[k] &= (\prod^{0}_{\tau=k-1}F[\tau]) x[0] \nonumber\\
	&= P
	\begin{bmatrix}
	\begin{matrix}
	1 & k\\
	0 & 1
	\end{matrix} & \mathbf{0}_{2\times(2N-2)} \\
	\mathbf{0}_{(2N-2)\times 2} & (\prod\limits^{0}_{\tau=k-1} \tilde{Q}[\tau]) 
	\end{bmatrix}
	P^{-1} {x}[0].
	\end{align*}
With $\lim_{k\to\infty}\prod\limits^{0}_{\tau=k-1} \tilde{Q}[\tau] =\mathbf{0}_{(2N-2)\times (2N-2)}$, one has 	
\begin{align*}
	& \lim_{k\to \infty} ( x[k] - P
	\begin{bmatrix}
	\begin{matrix}
	1 & k\\
	0 & 1
	\end{matrix} & \mathbf{0}_{2\times (2N-2)} \\
	\mathbf{0}_{  (2N-2)\times 2} & \mathbf{0}_{(2N-2)\times(2N-2)}
	\end{bmatrix}
	P^{-1} x[0] )\\
	= & 	\lim_{k\to \infty} ( x[k]- 
	\begin{bmatrix}
	\mathbf{1}_N  \w^\mathsf{T} & k \mathbf{1}_N  \w^\mathsf{T} \\
	\mathbf{0}_N & \mathbf{1}_N  \w^\mathsf{T} \\
	\end{bmatrix}
	\begin{bmatrix}
	p[0] \\ v[0]
	\end{bmatrix} ) =0,
\end{align*} which is equivalent to \eqref{consensus} with
\begin{align*}	
\sigma[k] = \left[ \begin{matrix}  \w^\mathsf{T}  p[0] + k \w^\mathsf{T} v[0] \\
 \w^\mathsf{T}   v[0] \end{matrix} \right].
\end{align*}
The proof is thus completed.
\eproof

\begin{figure}[t]
	\centering
	\hspace{-8mm}
	\includegraphics[scale=.46]{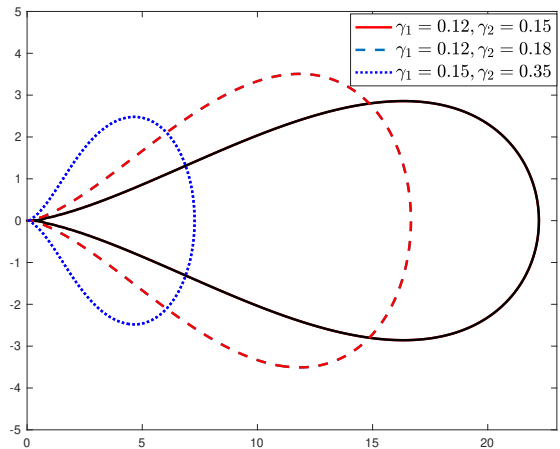}
	\caption{Boundaries of Laplacian eigenvalue for achieving structural consensus.}\label{fig-BoundL}
\end{figure}

\begin{remark}	 \label{remark-bound}
For MASs of second order dynamics, the conditions for $\gamma_1$ and $\gamma_2$ 
to guarantee structural consensus is expressed in a more complicated form than those for  first order dynamics.
The idea is to keep all the eigenvalues of $F[k]$ staying within the unit circle except 
for the eigenvalue $1$.
From the proof of Lemma~\ref{lemma:parameter}, 
it suffices to select the parameters $\gamma_1$ and $\gamma_2$ to satisfy 
\eqref{gammacon1}, \eqref{gammacon2}, and \eqref{gammacon3}. In fact, 
\eqref{gammacon2} is redundant and automatically implied by \eqref{gammacon1}.
In other words, with $\gamma_2 > \gamma_1$   fixed, the eigenvalues must satisfy 
\eqref{gammacon1}, which holds if they are located within the region bounded by the closed
curve defined by \eqref{curve} in a complex plane. 
With different choices of  $\gamma_1$ and $\gamma_2$, the closed curves are illustrated in
Fig.~\ref{fig-BoundL}.  It is worth mentioning that none of the Gershgorin disks of $L[k]$ can be inside this area, because all the disks are tangent to the imaginary axis.

\end{remark}

\section{Cryptographic Implementation}\label{sec-cryptography}

In a distributed network, communication happens between each pair of neighbors. Specifically, the control input $u_i[k]$ of each agent $i$ can be regarded as
\begin{equation}\label{equ-necessary_mesage}
	u_i[k]=\sum_{j\in \mathbb{N}_i}u_{ij}[k],
\end{equation}
where  $u_{ij}[k]$ is the information that agent $i$ requires from agent $j$, 
specified according to the consensus law \eqref{equ-model-input_1} or \eqref{equ-model-input_2}.
In the following discussion, $u_{ij}[k]$ is called  the ``message'' agent $i$ receives from agent $j$. 
If agent $i$ becomes a malicious neighbor of agent $j$, it intends to use the received messages to infer agent $j$'s initial condition, i.e., privacy.

In the following, we present a privacy-preserving algorithm for the agents to update their states
using Paillier algorithm \cite{paillier_public-key_1999}.
There exist two  set of keys, namely public key $K^{Pub}$ and private key $K^{Prv}$,  used for encryption and decryption, respectively. The algorithm includes generating keys, encrypting data, and decrypting data.
Paillier encryption has an additive semi-homomorphic property. Under Paillier encryption, the   property
$\mathcal{E}(m_1)\cdot\mathcal{E}(m_2)=\mathcal{E}(m_1+m_2)$ holds, 
where $m_1$ and $m_2$ are plaintexts and $\mathcal{E}$ denotes the encryption function.

We consider the scenario where one of the agents, say Alice,  updates $x_A[k]$ using the communication with Bob, one of its neighbors, without loss of generality.
  In order to protect the privacy, our algorithms consider injecting uncertainty into coupling couples, i.e., $a_{ij}[k]$s.
For simplicity of notation, in the algorithm, $x_A[k]$, $x_B[k]$, and $a_{AB}[k]$ are written as $x_A$, $x_B$, and $a_{AB}$, respectively. 
The term $\mathcal{E}_i(m)$ stands for the ciphertext that is computed by encrypting the plaintext $m$ with agent $i$'s public key $K_i^{Pub}$.

We consider Alice as a curious agent trying to calculate states of Bob, which is one of Alice's neighbors, by collecting its own states, i.e., 
$x_A[k]$  and the message received from Bob, i.e., $u_{AB}[k]$.
From time $k=0$ to $k=k_c$, the information obtained by Alice is listed as follows,
\begin{align}\label{equ-MessageCollecting-1}
u_{AB}[k]= & \epsilon a_{AB}[k](x_B[k]-x_A[k]),
\end{align}
  for the first order case, or, 
\begin{align}\label{equ-MessageCollecting-2}
u_{AB}[k]= & \gamma_1 a_{AB}[k](p_B[k]-p_A[k]) \nonumber\\
&+\gamma_2 a_{AB}[k](v_B[k]-v_A[k]),
\end{align}
for the second order case.
We see that in each iteration, Alice collects more unknowns than equations, which makes the equation set unsolvable. 
Hence privacy of both Alice and Bob is preserved under  Algorithms~\ref{alg-Directed_FirstOrder} and \ref{alg-Directed-SecondOrder}, when they adopts the first order and second order dynamics, respectively.

\begin{algorithm}
	\caption{Information exchange in directed networks (first order dynamics)}\label{alg-Directed_FirstOrder}
	\textbf{Preparation} (Alice):\\
	At initial time $k=0$, generate a pair of public key $K_A^{Pub}$ and private key $K_A^{Prv}$, then send $K_A^{Pub}$ to all its neighbors, including Bob. \\
	\textbf{Preparation} (Bob):\\
	 At time $k$, generate a random number within a certain interval:
	 $a_{AB}\in(a_{AB}[0]-\delta,a_{AB}[0]+\delta)$.\\
	\textbf{Step 1} (Alice):\\
	Encrypt state:\\
	 $x_A\rightarrow-x_A\xrightarrow{\mathcal E}\mathcal{E}_A(-x_A)\xrightarrow{sent\;to}$ Bob.
	\\
	\textbf{Step 2} (Bob):\\
	Operate state:\\
	 $x_B\xrightarrow{\mathcal E}\mathcal{E}_A(x_B)\rightarrow\mathcal{E}_A(x_B)\cdot\mathcal{E}_A(-x_A)=\mathcal{E}_A(x_B-x_A)\rightarrow(\mathcal{E}_A(x_B-x_A))^{ a_B}=\mathcal{E}_A(a_{AB}\cdot(x_B-x_A))\xrightarrow{sent\;to}$ Alice.
	\\
	\textbf{Step 3} (Alice):
	\\
	Decrypt state: \\
	$\mathcal{E}_A(a_{AB}\cdot(x_B-x_A))\xrightarrow{\mathcal{E}^{-1}} a_{AB}\cdot(x_B-x_A)=u_{AB}$.
\end{algorithm}

\begin{algorithm}
	\caption{Information exchange in directed networks (second order dynamics)}
	\label{alg-Directed-SecondOrder}
	\textbf{Preparation} (Alice):\\
	(1) At initial time $k=0$, generate a pair of public key $K_A^{Pub}$ and private key $K_A^{Prv}$, then send $K_A^{Pub}$ to all its neighbors, including Bob. \\
	\textbf{Preparation} (Bob):\\
	(1) At time $k$, generate a random number within a certain interval:
	$a_{AB}\in(a_{AB}[0]-\delta,a_{AB}[0]+\delta)$.\\
	\textbf{Step 1} (Alice):\\
	(1.1) Encrypt position: $p_A\rightarrow-p_A\xrightarrow{\mathcal E}\mathcal{E}_A(-p_A)\xrightarrow{sent\;to}$ Bob.
	\\
	(1.2) Encrypt velocity: $v_A\rightarrow-v_A\xrightarrow{\mathcal E}\mathcal{E}_A(-v_A)\xrightarrow{sent\;to}$ Bob.
	\textbf{Step 2} (Bob):\\
	(2.1) Operate position: $p_B\xrightarrow{\mathcal E}\mathcal{E}_A(p_B)\rightarrow\mathcal{E}_A(p_B)\cdot\mathcal{E}_A(-p_A)=\mathcal{E}_A(p_B-p_A)\rightarrow(\mathcal{E}_A(p_B-p_A))^{\gamma_1 a_{AB}}=\mathcal{E}_A(\gamma_1 a_{AB}\cdot(p_B-p_A))$.
	\\
	(2.2) Operate velocity: $v_B\xrightarrow{\mathcal E}\mathcal{E}_A(v_B)\rightarrow\mathcal{E}_A(v_B)\cdot\mathcal{E}_A(-v_A)=\mathcal{E}_A(v_B-v_A)\rightarrow(\mathcal{E}_A(v_B-v_A))^{\gamma_2 a_{AB}}=\mathcal{E}_A(\gamma_2 a_{AB}\cdot(v_B-v_A))$.
	\\
	(2.3) Combine $p$ and $v$:
	$\mathcal{E}_A(\gamma_1 a_{AB}\cdot(p_B-p_A))\cdot\mathcal{E}_A(\gamma_2 a_{AB}\cdot(v_B-v_A))=\mathcal{E}_A(\gamma_1 a_{AB}\cdot(p_B-p_A)+\gamma_2 a_{AB}\cdot(v_B-v_A))\xrightarrow{sent\;to}$ Alice. 
	\\
	\textbf{Step 3} (Alice):
	\\
	Decrypt and operate:\\
	$\mathcal{E}_A(\gamma_1 a_{AB}\cdot(p_B-p_A)+\gamma_2 a_{AB}\cdot(v_B-v_A))\xrightarrow{\mathcal{E}^{-1}} \gamma_1 a_{AB}\cdot(p_B-p_A)+\gamma_2 a_{AB}\cdot(v_B-v_A)=u_{AB}$.
\end{algorithm}

The statement regarding the second objective is summarized in the following theorems for first order dynamics
and second order dynamics, respectively. 

\begin{theorem}\label{lem-privacy1}
Consider the MAS \eqref{equ-SystemModel_G} with \eqref{F1} implemented in Algorithm~\ref{alg-Directed_FirstOrder}
in a network equipped with a directed leader-following topology containing a spanning tree. 
For a specific $\delta$ satisfying \eqref{delta}, $a_{ij}[k]$'s satisfying \eqref{aij} are assigned and known only
by agent $j$, and the parameter $\epsilon$ satisfies \eqref{con-stable}. 
The MAS achieves structural consensus and all the agents' privacy are preserved.
\end{theorem}

\begin{theorem}\label{lem-privacy2}
Consider the MAS \eqref{equ-SystemModel_G} with \eqref{F2} implemented in Algorithm~\ref{alg-Directed-SecondOrder}
in a network equipped with a directed leader-following topology containing a spanning tree. 
For a specific $\delta$ satisfying \eqref{delta}, $a_{ij}[k]$'s satisfying \eqref{aij} are assigned and known only
by agent $j$, and the parameters $\gamma_1$ and $\gamma_2$ are selected in Lemmas~\ref{lemma:eigenvalue} and 
\ref{lemma:parameter}.
The MAS achieves structural consensus and all the agents' privacy are preserved.
\end{theorem}

\noindent{\it Proof:} The proofs for the two theorems are similar and only that for Theorem~\ref{lem-privacy1} is 
given below. 
It is known from the previous section that the MAS achieves structural consensus. So, we only 
prove the preservation of privacy.  It suffices to examine the 
possibility of agent's information being disclosed to either an eavesdropper or a malicious neighbor. 
Let us consider an arbitrary pair of agents, say Alice and Bob, and check if Alice or Bob's privacy is violated during the process of updating Alice's state $x_A[k]$.
	
 Firstly, an eavesdropper can obtain access to the communicating messages. 
In particular,  in Steps 1 and 2 of Algorithm \ref{alg-Directed_FirstOrder}, the messages are $\mathcal{E}_A(-x_A)$ and $\mathcal{E}_A(a_{AB}\cdot(x_B-x_A))$. Since these messages are encrypted by Alice's public key $K_A^{Pub}$, they can not be decrypted by an eavesdropper without Alice's private key $K_A^{Pri}$. Therefore,
neither Alice nor Bob's state information is disclosed.

Secondly, we examine if Bob (or Alice), when regarded as a malicious agent, 
can infer the privacy of Alice (or Bob) using the available information. 
According to Algorithm \ref{alg-Directed_FirstOrder}, Bob receives only one message, i.e., $\mathcal{E}_A(-x_A)$, which is incomprehensible because he does not have $K_A^{Pub}$.
So, when Bob is a malicious neighbor, he can not infer the privacy of Alice. 
Next, we consider the case where Alice is a malicious neighbor and she 
aims to reconstruct Bob's state by collecting the messages received from Bob, i.e., $u_{AB}[k]$, as well as its own states $x_A[k]$.
	From $k=0$ to $k=k_c$, the information obtained by Alice is listed as follows
	\begin{equation}\label{equ-MessageCollecting-1}
	u_{AB}[k]=a_{AB}[k](x_B[k]-x_A[k]),\quad k=0,1,\dots k_c,
	\end{equation}
	where $u_{AB}[k]$ and $x_A[k]$ are known to Alice. From the set of equations \eqref{equ-MessageCollecting-1}, we see that in each iteration, Alice collects more unknowns than equations, which makes the set of equations unsolvable for a unique solution. Therefore, Alice can not infer the privacy of Bob. 
	 \eproof

\begin{remark}\label{rmk-weights_constant}
It is worth mentioning that 
time-varying weights $a_{ij}[k]$'s are used in the networked system under consideration. 
This is because constant weights may result in higher risk of disclosure of agents' trajectories. 
Suppose Alice tends to infer Bob's states by collecting the messages $u_{AB}[k]$
with a constant $a_{AB}$.  As a result, she can generate a set of equations as follows
		\begin{align}
			u_{AB}[0]=&a_{AB}(x_B[0]-x_A[0]),\nonumber\\
			&\vdots \nonumber\\
			u_{AB}[k]=&a_{AB}(x_B[k]-x_A[k]), \label{equ-constant_weights}
		\end{align}
which contains significantly less unknown variables.
Consider a possible scenario in which $\frac{u_{AB}[T]}{u_{AB}[0]}=\frac{x_A[T]}{x_A[0]}=\beta$, one can easily conclude that the Bob's position is $x_B[T]=\beta x_B[0]$.  In other words, Alice is able to infer Bob's initial state (privacy) 
at time $T$.


	
\end{remark}

\begin{remark} For a networked system of an undirected topology, it is impossible 
to keep the weight $a_{ij}$ to agent $j$  because of the symmetric property 
$a_{ij} =a_{ji}$. Therefore,  Algorithm~\ref{alg-Directed_FirstOrder} or \ref{alg-Directed-SecondOrder}
does not apply for undirected topologies.  
Different design strategies for $a_{ij}$ can be found in \cite{ruan_secure_2017, fang_secure_2018} for undirected topologies. 
Also,  for a networked system of an undirected topology, one has 
$u_{ij} = - u_{ji}$, regardless of the design strategy of  $a_{ij}$. 
When consensus occurs at $k$, Alice knows Bob's state from $x_B[k]=x_A[k]$
and his input $u_B[k] = u_{BA}[k] = - u_{AB}[k]$ when Bob has a sole neighbor Alice. 
In this scenario, Alice can infer Bob's initial state using the dynamics $x_B[k+1]=x_B[k] + u_B[k]$. 
However, for the directed topology case, this risk is avoided because a malicious agent can not 
take advantage of the property $u_{ij} = - u_{ji}$. 
\end{remark}

\section{Simulation}\label{sec-simulation}
 
\begin{figure}[t]
	\centering
	\includegraphics[scale=.35]{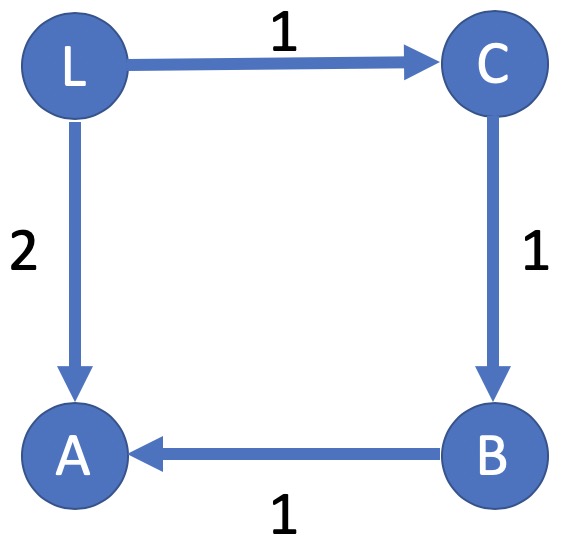}
	\caption{A directed network of four agents. }\label{fig-topology}
\end{figure}

In Section \ref{sec-cryptography}, it is shown that the agents' states can not be derived by knowing system dynamics as well as collecting communicating messages.
In this section, numerical simulation is used to further demonstrate the established result.
We consider a leader-following system consisting of  four agents, namely $L$, $A$, $B$ and $C$.
The topology presented in Fig \ref{fig-topology} with the initial coupling weights set as $1$.
 In particular, the agent $A$ computes $u_A[k]$ to update its state, i.e., $x_A[k]$, by communicating with $B$ and $L$. 
We present the communicating messages between agents $A$ and $B$ to show that their states  cannot be derived by collecting messages.

\begin{figure}[htb]
	\centering
	\includegraphics[scale=.4]{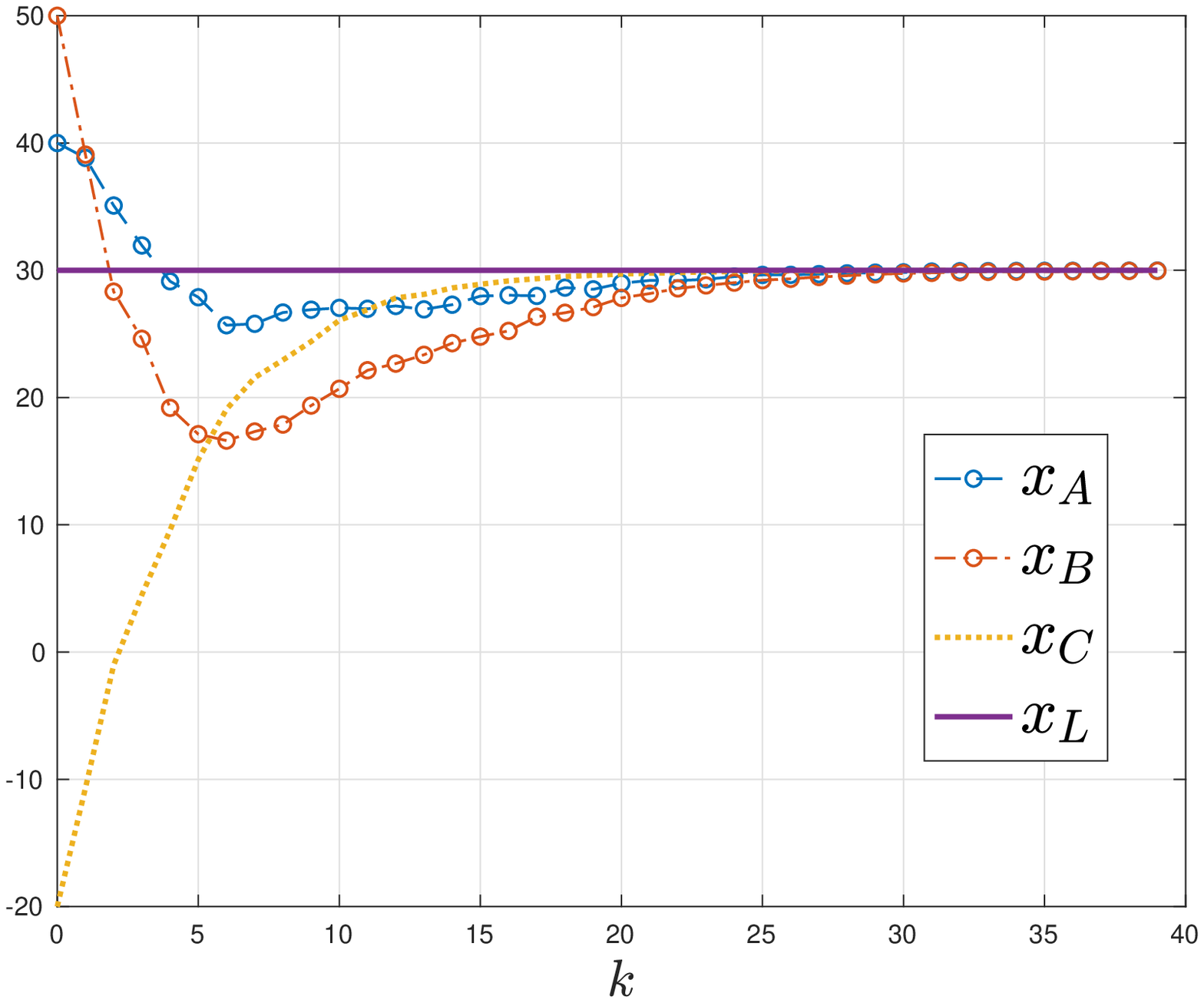}
	\caption{Trajectories of agents' states in the simulated network.}\label{fig-consensus}
%
	\centering
	\includegraphics[scale=.4]{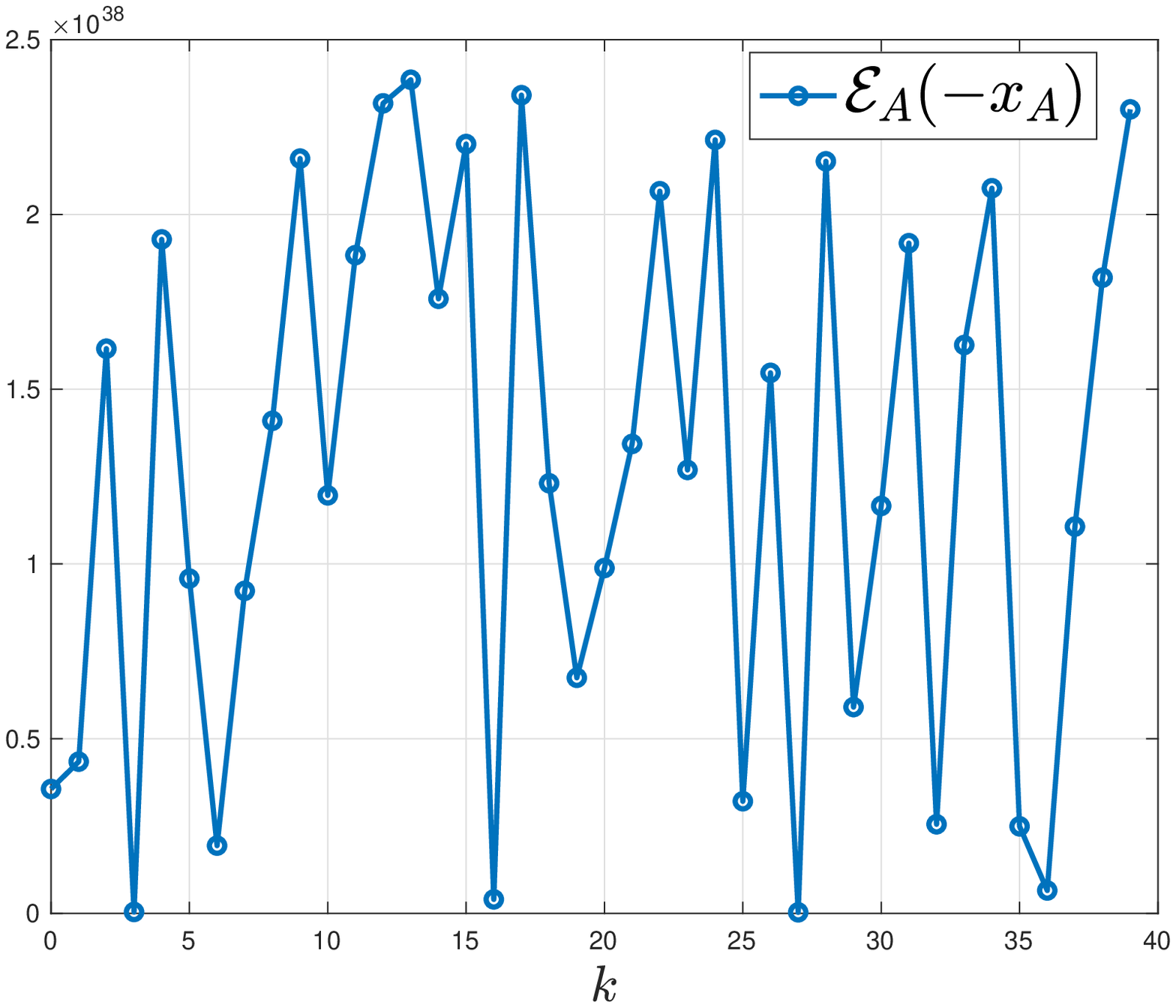}
	\caption{Message sent by agent A in Step~1 of Algorithm~\ref{alg-Directed_FirstOrder}. }\label{fig-exa}
 \end{figure}
 
\begin{figure}[htb]
     \centering
	\centering
	\includegraphics[scale=.4]{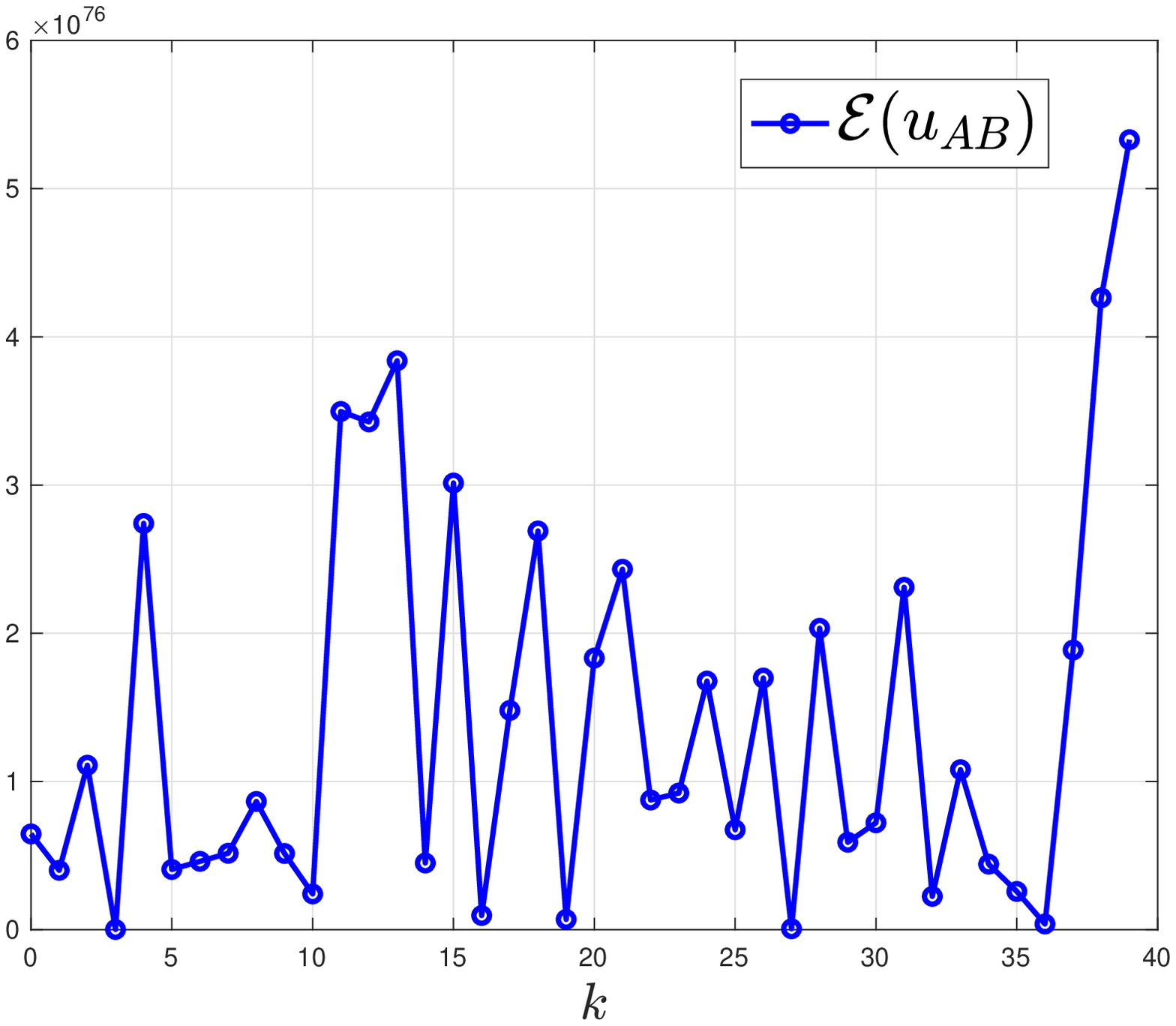}
	\caption{Message sent by agent B in Step~2 of Algorithm~\ref{alg-Directed_FirstOrder}. }\label{fig-euab}
%
%
   \centering
	\centering
	\includegraphics[scale=.4]{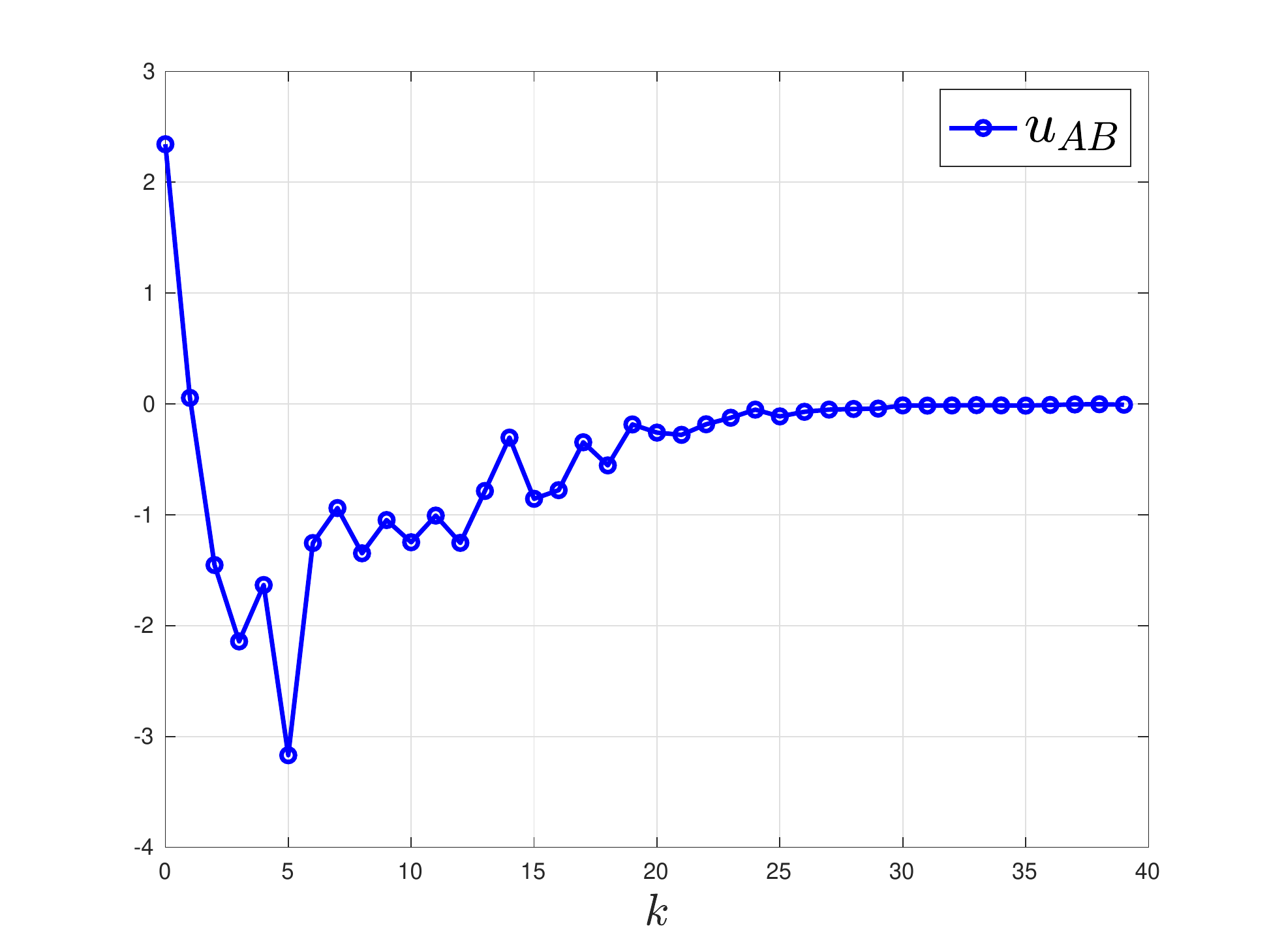}
	\caption{Message received and encrypted by agent A  in Step~3 of Algorithm~\ref{alg-Directed_FirstOrder}. }\label{fig-uab}
\end{figure}
The first simulation was conducted for the agents of 
first order dynamics  with the initial states $x_L[0]=30$, $x_A[0]=40$, $x_B[0]=50$, and $x_C[0]=-20$.
The specific $\delta=0.5$ satisfying \eqref{delta} is selected and 
the control parameter is $\epsilon=0.4$.
As shown in Fig. \ref{fig-consensus}, the actual states of agents  $A$, $B$ and $C$ converge to $x_L[k]$, which is the constant state of the leader $L$. From Fig.~\ref{fig-exa} and Fig.~\ref{fig-euab}, one can observe that without decryption, the messages transmitted in the network are incomprehensible.
The actually input  received and encrypted by agent A is shown in Fig.~\ref{fig-uab}.
In addition, the average computational time cost for each agent at each iteration
in Algorithm \ref{alg-Directed_FirstOrder} is presented in Table \ref{tab-ComputingTime},
 using Matlab R2018b on a CPU of $2.3$ GHz Quad-Core Intel Core i$5$.

 \begin{figure}[htb]
	\centering
	\includegraphics[scale=.46]{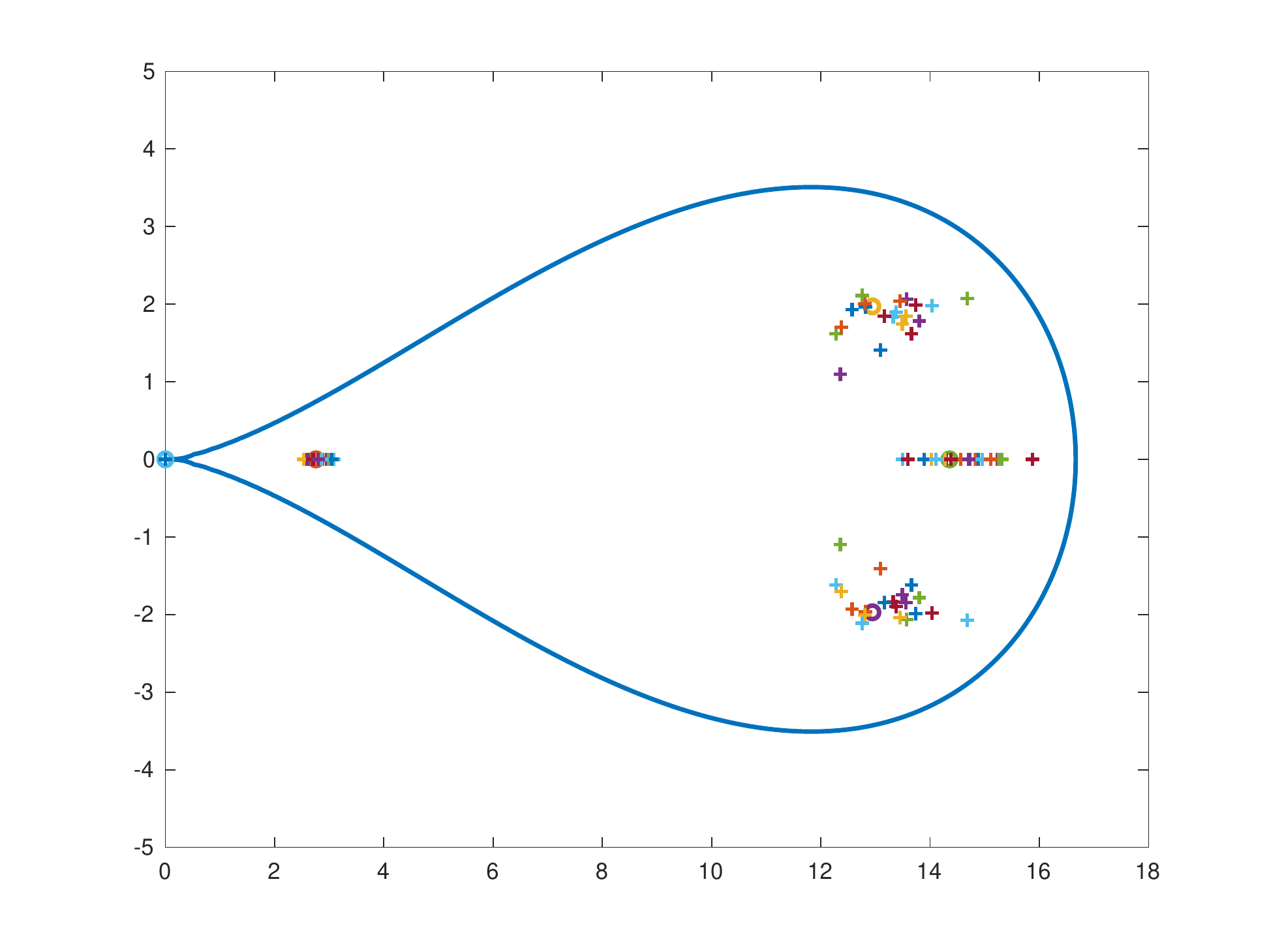}
	\caption{Location of nonzero eigenvalues of $L[k]$ inside the closed boundary specified in Remark~\ref{remark-bound}. }\label{fig-SecondOrder}
\end{figure}

\begin{table}[ht] 		
\caption{Time cost of different processes in Algorithm \ref{alg-Directed_FirstOrder}.}	 
\centering
	\begin{tabular}{|p{3cm}|p{1cm}|p{1cm}|p{1cm}|}
		\hline
		Key bits & $32$  &  $64$ & $128$ \\
		\hline
		Encryption (ms) &  $1.1442$ & $1.2461$ & $1.3032$\\
		\hline
		Controller (ms) & $0.0752$  &  $0.0182$ & $0.0198$ \\
		\hline
		Decryption (ms) &  $0.0458$ & $0.0166$ & $0.0343$\\
		\hline\end{tabular}
	\label{tab-ComputingTime}
\end{table}

The simulation was repeated for five agents of second order dynamics
by applying Algorithm~\ref{alg-Directed-SecondOrder} with the parameters $\delta=0.5$, 
$\gamma_1=0.3$, and $\gamma_2=0.75$.
Similar results were observed and the plots are omitted here. 
It is worth noting that the nonzero eigenvalues of the Laplacian $L[k]$
are located  within the region bounded by the closed
curve defined by \eqref{curve} in a complex plane, as illustrated in Fig.~\ref{fig-SecondOrder}.

 \section{Conclusion}\label{sec-conclusion}

In this paper, a Paillier encryption based privacy-preserving consensus algorithm 
	has been presented to prevent agents' initial states from being disclosed to
	either an eavesdropper or a malicious neighbor.
	We have studied a directed leader-following topology for both first order and second order systems.
 	Given that information is sent under encryption with the public key of an agent, it remains intact against eavesdroppers.
		We have also shown that the privacy of agents would not be violated by malicious neighbors.
	Together with the proof of confidentiality, we have also presented the sufficient conditions for reaching structural consensus under a directed network with time-varying coupling weights.
	
 \bibliographystyle{model1-num-names}

 \bibliography{PrivateConsensus}%

\begin{thebibliography}{25}
\expandafter\ifx\csname natexlab\endcsname\relax\def\natexlab#1{#1}\fi
\providecommand{\bibinfo}[2]{#2}
\ifx\xfnm\relax \def\xfnm[#1]{\unskip,\space#1}\fi
\bibitem[{{Olfati-Saber} and Murray(2004)}]{olfati-saber_consensus_2004}
\bibinfo{author}{R.~{Olfati-Saber}}, \bibinfo{author}{R.~M. Murray},
\newblock \bibinfo{title}{Consensus {{Problems}} in {{Networks}} of {{Agents}}
  with {{Switching Topology}} and {{Time}}-delays},
\newblock \bibinfo{journal}{IEEE Transactions on Automatic Control}
  \bibinfo{volume}{49} (\bibinfo{year}{2004}) \bibinfo{pages}{1520--1533}.
\bibitem[{Jadbabaie et~al.(2003)Jadbabaie, Lin, and
  Morse}]{jadbabaie_coordination_2003}
\bibinfo{author}{A.~Jadbabaie}, \bibinfo{author}{J.~Lin},
  \bibinfo{author}{A.~S. Morse},
\newblock \bibinfo{title}{Coordination of {{Groups}} of {{Mobile Autonomous
  Agents Using Nearest Neighbor Rules}}},
\newblock \bibinfo{journal}{IEEE Transactions on Automatic Control}
  \bibinfo{volume}{48} (\bibinfo{year}{2003}) \bibinfo{pages}{988--1001}.
\bibitem[{Xie and Wang(2012)}]{xie_consensus_2012}
\bibinfo{author}{D.~Xie}, \bibinfo{author}{S.~Wang},
\newblock \bibinfo{title}{Consensus of {{Second}}-order {{Discrete}}-time
  {{Multi}}-agent {{Systems}} with {{Fixed Topology}}},
\newblock \bibinfo{journal}{Journal of Mathematical Analysis and Applications}
  \bibinfo{volume}{387} (\bibinfo{year}{2012}) \bibinfo{pages}{8--16}.
\bibitem[{Lin and Jia(2009)}]{lin_consensus_2009}
\bibinfo{author}{P.~Lin}, \bibinfo{author}{Y.~Jia},
\newblock \bibinfo{title}{Consensus of {{Second}}-order {{Discrete}}-time
  {{Multi}}-agent {{Systems}} with {{Nonuniform Time}}-delays and {{Dynamically
  Changing Topologies}}},
\newblock \bibinfo{journal}{Automatica} \bibinfo{volume}{45}
  (\bibinfo{year}{2009}) \bibinfo{pages}{2154--2158}.
\bibitem[{Yu et~al.(2010)Yu, Chen, and Cao}]{yu_necessary_2010}
\bibinfo{author}{W.~Yu}, \bibinfo{author}{G.~Chen}, \bibinfo{author}{M.~Cao},
\newblock \bibinfo{title}{Some necessary and sufficient conditions for
  second-order consensus in multi-agent dynamical systems},
\newblock \bibinfo{journal}{Automatica} \bibinfo{volume}{46}
  (\bibinfo{year}{2010}) \bibinfo{pages}{1089--1095}.
\bibitem[{Yu et~al.(2011)Yu, Chen, Ren, Kurths, and
  Zheng}]{yu_distributed_2011}
\bibinfo{author}{W.~Yu}, \bibinfo{author}{G.~Chen}, \bibinfo{author}{W.~Ren},
  \bibinfo{author}{J.~Kurths}, \bibinfo{author}{W.~X. Zheng},
\newblock \bibinfo{title}{Distributed {{Higher Order Consensus Protocols}} in
  {{Multiagent Dynamical Systems}}},
\newblock \bibinfo{journal}{IEEE Transactions on Circuits and Systems I:
  Regular Papers} \bibinfo{volume}{58} (\bibinfo{year}{2011})
  \bibinfo{pages}{1924--1932}.
\bibitem[{Zhu and Mart{\'i}nez(2010)}]{zhu_discrete-time_2010}
\bibinfo{author}{M.~Zhu}, \bibinfo{author}{S.~Mart{\'i}nez},
\newblock \bibinfo{title}{Discrete-time dynamic average consensus},
\newblock \bibinfo{journal}{Automatica} \bibinfo{volume}{46}
  (\bibinfo{year}{2010}) \bibinfo{pages}{322--329}.
\bibitem[{Kingston and Beard(2006)}]{kingston_discrete-time_2006}
\bibinfo{author}{D.~B. Kingston}, \bibinfo{author}{R.~W. Beard},
\newblock \bibinfo{title}{Discrete-time average-consensus under switching
  network topologies},
\newblock in: \bibinfo{booktitle}{2006 {{American Control Conference}}}, pp.
  \bibinfo{pages}{6 pp.--}.
\bibitem[{LeBlanc and Koutsoukos(2011)}]{leblanc_consensus_2011}
\bibinfo{author}{H.~J. LeBlanc}, \bibinfo{author}{X.~D. Koutsoukos},
\newblock \bibinfo{title}{Consensus in {{Networked Multi}}-agent {{Systems}}
  with {{Adversaries}}},
\newblock in: \bibinfo{booktitle}{Proceedings of the 14th {{International
  Conference}} on {{Hybrid Systems}}: {{Computation}} and {{Control}}},
  {{HSCC}} '11, \bibinfo{publisher}{{ACM}}, \bibinfo{address}{{New York, NY,
  USA}}, \bibinfo{year}{2011}, pp. \bibinfo{pages}{281--290}.
\bibitem[{LeBlanc et~al.(2012)LeBlanc, Zhang, Sundaram, and
  Koutsoukos}]{leblanc_consensus_2012}
\bibinfo{author}{H.~J. LeBlanc}, \bibinfo{author}{H.~Zhang},
  \bibinfo{author}{S.~Sundaram}, \bibinfo{author}{X.~Koutsoukos},
\newblock \bibinfo{title}{Consensus of {{Multi}}-agent {{Networks}} in the
  {{Presence}} of {{Adversaries Using Only Local Information}}},
\newblock in: \bibinfo{booktitle}{Proceedings of the 1st {{International
  Conference}} on {{High Confidence Networked Systems}}}, {{HiCoNS}} '12,
  \bibinfo{publisher}{{ACM}}, \bibinfo{address}{{New York, NY, USA}},
  \bibinfo{year}{2012}, pp. \bibinfo{pages}{1--10}.
\bibitem[{Feng et~al.(2016)Feng, Hu, and Wen}]{feng_distributed_2016}
\bibinfo{author}{Z.~Feng}, \bibinfo{author}{G.~Hu}, \bibinfo{author}{G.~Wen},
\newblock \bibinfo{title}{Distributed consensus tracking for multi-agent
  systems under two types of attacks},
\newblock \bibinfo{journal}{International Journal of Robust and Nonlinear
  Control} \bibinfo{volume}{26} (\bibinfo{year}{2016})
  \bibinfo{pages}{896--918}.
\bibitem[{Manitara and Hadjicostis(2013)}]{manitara_privacy-preserving_2013}
\bibinfo{author}{N.~E. Manitara}, \bibinfo{author}{C.~N. Hadjicostis},
\newblock \bibinfo{title}{Privacy-preserving {{Asymptotic Average Consensus}}},
\newblock in: \bibinfo{booktitle}{2013 {{European Control Conference}}}, pp.
  \bibinfo{pages}{760--765}.
\bibitem[{Mo and Murray(2017)}]{mo_privacy_2017}
\bibinfo{author}{Y.~Mo}, \bibinfo{author}{R.~M. Murray},
\newblock \bibinfo{title}{Privacy {{Preserving Average Consensus}}},
\newblock \bibinfo{journal}{IEEE Transactions on Automatic Control}
  \bibinfo{volume}{62} (\bibinfo{year}{2017}) \bibinfo{pages}{753--765}.
\bibitem[{Duan et~al.(2015)Duan, He, Cheng, Mo, and Chen}]{duan_privacy_2015}
\bibinfo{author}{X.~Duan}, \bibinfo{author}{J.~He}, \bibinfo{author}{P.~Cheng},
  \bibinfo{author}{Y.~Mo}, \bibinfo{author}{J.~Chen},
\newblock \bibinfo{title}{Privacy {{Preserving Maximum Consensus}}},
\newblock in: \bibinfo{booktitle}{2015 54th {{IEEE Conference}} on {{Decision}}
  and {{Control}}}, pp. \bibinfo{pages}{4517--4522}.
\bibitem[{Ruan et~al.(2017)Ruan, Ahmad, and Wang}]{ruan_secure_2017}
\bibinfo{author}{M.~Ruan}, \bibinfo{author}{M.~Ahmad},
  \bibinfo{author}{Y.~Wang},
\newblock \bibinfo{title}{Secure and {{Privacy}}-{{Preserving Average
  Consensus}}},
\newblock in: \bibinfo{booktitle}{Proceedings of the 2017 {{Workshop}} on
  {{Cyber}}-{{Physical Systems Security}} and {{Privacy}}}, {{CPS}} '17,
  \bibinfo{address}{{New York, USA}}, pp. \bibinfo{pages}{123--129}.
\bibitem[{Fang et~al.(2018)Fang, Zamani, and Chen}]{fang_secure_2018}
\bibinfo{author}{W.~Fang}, \bibinfo{author}{M.~Zamani},
  \bibinfo{author}{Z.~Chen},
\newblock \bibinfo{title}{Secure and {{Privacy Preserving Consensus}} for
  {{Second}}-order {{Systems Based}} on {{Paillier Encryption}}},
\newblock \bibinfo{journal}{arXiv:1805.01065 [cs]}  (\bibinfo{year}{2018}).
\bibitem[{Krutz and Vines(2010)}]{krutz_cloud_2010}
\bibinfo{author}{R.~L. Krutz}, \bibinfo{author}{R.~D. Vines},
  \bibinfo{title}{Cloud {{Security}}: {{A Comprehensive Guide}} to {{Secure
  Cloud Computing}}}, \bibinfo{publisher}{{Wiley Publishing}},
  \bibinfo{year}{2010}.
\bibitem[{Farokhi et~al.(2016)Farokhi, Shames, and
  Batterham}]{farokhi_secure_2016}
\bibinfo{author}{F.~Farokhi}, \bibinfo{author}{I.~Shames},
  \bibinfo{author}{N.~Batterham},
\newblock \bibinfo{title}{Secure and {{Private Cloud}}-{{Based Control Using
  Semi}}-{{Homomorphic Encryption}}},
\newblock \bibinfo{journal}{IFAC-PapersOnLine} \bibinfo{volume}{49}
  (\bibinfo{year}{2016}) \bibinfo{pages}{163--168}.
\bibitem[{Sadeghikhorami et~al.(2020)Sadeghikhorami, Zamani, Chen, and
  Safavi}]{sadeghikhorami2020secure}
\bibinfo{author}{L.~Sadeghikhorami}, \bibinfo{author}{M.~Zamani},
  \bibinfo{author}{Z.~Chen}, \bibinfo{author}{A.~A. Safavi},
\newblock \bibinfo{title}{A secure control mechanism for network environments},
\newblock \bibinfo{journal}{Journal of the Franklin Institute}
  \bibinfo{volume}{357} (\bibinfo{year}{2020}) \bibinfo{pages}{12264--12280}.
\bibitem[{Ruan et~al.(2019)Ruan, Gao, and Wang}]{ruan_secure_2019-1}
\bibinfo{author}{M.~Ruan}, \bibinfo{author}{H.~Gao}, \bibinfo{author}{Y.~Wang},
\newblock \bibinfo{title}{Secure and {{Privacy}}-{{Preserving Consensus}}},
\newblock \bibinfo{journal}{IEEE Transactions on Automatic Control}
  \bibinfo{volume}{64} (\bibinfo{year}{2019}) \bibinfo{pages}{4035--4049}.
\bibitem[{He et~al.(2019)He, Cai, Zhao, Cheng, and
  Guan}]{he_privacy-preserving_2019}
\bibinfo{author}{J.~He}, \bibinfo{author}{L.~Cai}, \bibinfo{author}{C.~Zhao},
  \bibinfo{author}{P.~Cheng}, \bibinfo{author}{X.~Guan},
\newblock \bibinfo{title}{Privacy-{{Preserving Average Consensus}}: {{Privacy
  Analysis}} and {{Algorithm Design}}},
\newblock \bibinfo{journal}{IEEE Transactions on Signal and Information
  Processing over Networks} \bibinfo{volume}{5} (\bibinfo{year}{2019})
  \bibinfo{pages}{127--138}.
\bibitem[{Liu et~al.(2017)Liu, Ren, and Mo}]{liu_secure_2017}
\bibinfo{author}{Q.~Liu}, \bibinfo{author}{X.~Ren}, \bibinfo{author}{Y.~Mo},
\newblock \bibinfo{title}{Secure and privacy preserving average consensus},
\newblock in: \bibinfo{booktitle}{2017 11th {{Asian Control Conference}}
  ({{ASCC}})}, pp. \bibinfo{pages}{274--279}.
\bibitem[{{Ching-Tai Lin}(1974)}]{lin1974structural}
\bibinfo{author}{{Ching-Tai Lin}},
\newblock \bibinfo{title}{Structural controllability},
\newblock \bibinfo{journal}{IEEE Transactions on Automatic Control}
  \bibinfo{volume}{19} (\bibinfo{year}{1974}) \bibinfo{pages}{201--208}.
\bibitem[{Mehrabadi et~al.(2019)Mehrabadi, Zamani, and
  Chen}]{mehrabadi_structural_2019}
\bibinfo{author}{M.~K. Mehrabadi}, \bibinfo{author}{M.~Zamani},
  \bibinfo{author}{Z.~Chen},
\newblock \bibinfo{title}{Structural {{Controllability}} of a {{Consensus
  Network With Multiple Leaders}}},
\newblock \bibinfo{journal}{IEEE Transactions on Automatic Control}
  \bibinfo{volume}{64} (\bibinfo{year}{2019}) \bibinfo{pages}{5101--5107}.
\bibitem[{Paillier(1999)}]{paillier_public-key_1999}
\bibinfo{author}{P.~Paillier},
\newblock \bibinfo{title}{Public-{{Key Cryptosystems Based}} on {{Composite
  Degree Residuosity Classes}}},
\newblock in: \bibinfo{booktitle}{Advances in {{Cryptology}} \textemdash{}
  {{EUROCRYPT}} '99}, Lecture {{Notes}} in {{Computer Science}},
  \bibinfo{publisher}{{Springer, Berlin, Heidelberg}}, \bibinfo{year}{1999},
  pp. \bibinfo{pages}{223--238}.

\end{thebibliography}

\end{document}